\documentclass[aps,superscriptaddress,nofootinbib,twocolumn]{revtex4-1}
\usepackage{amsmath,latexsym,amssymb,hyperref,graphicx}
\usepackage{slashed}

\renewcommand\baselinestretch{1.02}

\hypersetup{colorlinks,citecolor= nicegreen,linkcolor= nicered}
\usepackage{color}
\definecolor{nicered}{rgb}{0.7,0.1,0.1}
\definecolor{nicegreen}{rgb}{0.1,0.5,0.1}

\newcommand\SEC[1]{\medskip\noindent{\sl\bfseries #1}}
\renewcommand{\figurename}{\renewcommand\baselinestretch{1.03}\em\footnotesize Fig.{}}
\renewcommand{\tablename}{\renewcommand\baselinestretch{1.03}\em\footnotesize  Table}

\newcommand{\beq}[1]{\begin{equation}\label{#1}}
  \newcommand{\eeq}[1]{\label{#1}\end{equation}}
\newcommand{\be}{\begin{equation}}
  \newcommand{\ee}{\end{equation}}
\newcommand{\bea}{\begin{eqnarray}}
  \newcommand{\eea}{\end{eqnarray}}

\newcommand{\beqn}[1]{\begin{eqnarray}\label{#1}}
  \newcommand{\eeqn}{\end{eqnarray}}
\newcommand{\bd}{\begin{displaymath}}
  \newcommand{\ed}{\end{displaymath}}
\newcommand{\mat}[4]{\left(\begin{array}{cc}{#1}&{#2}\\{#3}&{#4}
    \end{array}\right)}

\def\om{\omega}

\newcommand{\ov}{\overline}
\renewcommand{\to}{\rightarrow}
\renewcommand{\vec}[1]{\mathbf{#1}}

\renewcommand{\vec}[1]{ \boldsymbol{#1}}
\newcommand{\vect}[1]{\mbox{\boldmath$#1$}}

\newcommand{\dm}{\varepsilon}

\def\bsig{\mbox{\boldmath $\sigma$} }

\def\cD{{\cal D}}

\def\cN{{\mathcal N}}

\def\cP{\mathcal P}

\def\ff{{\rm f}}

\def\rT{\rm T}
\def\rt{\rm t}

\begin{document}

\title{Magnetic anomaly in UCN trapping: signal for neutron oscillations to parallel world?}

\author{Z. Berezhiani}
\affiliation{Dipartimento di Fisica, Universit\`a dell'Aquila, Via Vetoio, 67100 Coppito, L'Aquila, Italy} 
\affiliation{INFN, Laboratori Nazionali Gran Sasso, 67100 Assergi,  L'Aquila, Italy}

\author{F. Nesti}
\affiliation{Dipartimento di Fisica, Universit\`a dell'Aquila, Via Vetoio, 67100 Coppito, L'Aquila, Italy}

\date{\today}

\begin{abstract}
  \noindent 
  Present experiments do not exclude that the neutron transforms into some invisible degenerate
  twin, so called mirror neutron, with an appreciable probability.  These transitions are actively
  studied by monitoring neutron losses in ultra-cold neutron traps, where they can be revealed by
  their magnetic field dependence.
  In this work we reanalyze the experimental data acquired by the group of A.P. Serebrov at
  Institute Laue-Langevin, and find a dependence at more than $5\sigma$ away from the
  null hypothesis.
  This anomaly can be interpreted as oscillation to mirror neutrons with a timescale of few seconds,
  in the presence of a mirror magnetic field $B'\sim 0.1\,$G at the Earth.  If confirmed by future
  experiments, this will have a number of deepest consequences in particle physics and
  astrophysics.
\end{abstract}

\maketitle

\noindent 
There may exist a hidden parallel gauge sector that exactly copies the pattern of ordinary gauge
sector.  Then all particles: the electron $e$, proton $p$, neutron $n$ etc., should have invisible
twins: $e'$, $p'$, $n'$, etc.\ which are sterile to our strong and electroweak interactions
($SU(3)\times SU(2)\times U(1)$) but have their own gauge interactions ($SU(3)'\times SU(2)'\times
U(1)'$) with exactly the same couplings.  A notorious example, coined as mirror world~\cite{Mirror},
was introduced long time ago against parity violation: for our particles being left-handed, parity
can be interpreted as a discrete {\it mirror} symmetry which exchanges them with their twins which
are assumed to be right-handed.  Concerns about parity are irrelevant for our following discussions:
they extend to a parallel sector (or sectors) of any chirality. Nevertheless, in the following we
shall name the twin particles from the `primed' parallel sector as mirror particles.

Mirror matter can be a viable candidate for dark matter \cite{BCV}.  The baryon asymmetries in both
sectors can be generated by $B\!-\!L$ and $CP$ violating processes between ordinary and mirror
particles~\cite{BB-PRL}.  This scenario can naturally explain the relation 
$\Omega_D/\Omega_B \simeq 5$ 
between the dark and visible matter fractions in the Universe~\cite{IJMPA-B}.
The relevant interactions can be mediated by heavy messengers coupled to both sectors, as right
handed neutrinos~\cite{BB-PRL} or extra gauge bosons/gauginos~\cite{PLB98}.  In the context of extra
dimensions, ordinary and mirror sectors can be modeled as two parallel three-dimensional branes and
particle processes between them mediated by the bulk modes or ``baby branes" can be
envisaged~\cite{Gia}.

On the other hand, these interactions can induce mixing phenomena between ordinary and mirror
particles.  In fact, any {\it neutral} particle, elementary or composite, may oscillate into its
mirror twin, as e.g.\ ordinary neutrinos $\nu_e$, $\nu_\mu$, $\nu_\tau$ into their mirror partners,
sterile neutrinos $\nu'_e$, $\nu'_\mu$, $\nu'_\tau$~\cite{FV}.  A kinetic mixing between photon and
mirror photon~\cite{Holdom} would induce the positronium -- mirror positronium transition which is
searched for experimentally~\cite{Glashow}. Interestingly, this kinetic mixing may be responsible
also for the dark matter signals observed by the DAMA, CoGeNT and CRESST experiments~\cite{dama}.

As it was shown in ref.~\cite{BB-nn}, neither existing experimental limits nor cosmological and
astrophysical bounds can exclude the possibility that the oscillation between the neutron $n$ and
its mirror twin $n'$ is a rather fast process, hypothesis which can be tested in table-top
laboratory experiments.  The mass mixing, $\dm (\ov{n} n' + \ov{n}' n)$, can emerge from
$B$-violating six-fermion effective operators $(udd)(u'd'd')/M^5$ involving ordinary ($u,d$) and
mirror ($u',d'$) quarks, with $\dm \sim \Lambda_{\rm QCD}^6/M^5$ where $M$ is a cutoff scale of the
respective new physics and $\Lambda_{\rm QCD}\sim 250$~MeV is the scale of strong interactions.
Since the masses of $n$ and $n'$ are exactly equal, they have maximal mixing in vacuum and oscillate
with timescale $\tau = \dm^{-1}\sim (M/10\, {\rm TeV})^5~{\rm s}$. (In this paper we use
  natural units, $\hbar = c =1$.)  It is striking that present probes do not exclude $n$--$n'$
oscillation faster than the neutron decay, $\tau < \tau_n \approx 880\,$s.
The reason is that for neutrons bounded in nuclei, a $n$--$n'$ transition is forbidden by energy
conservation,  while $\tau \sim 1\,$s is compatible with the bounds
from primordial nucleosynthesis  and neutron star stability. 
As for free neutrons, oscillation is affected by magnetic fields and coherent interactions
with matter~\cite{BB-nn,EPJ}.

In ref.~\cite{BB-nn} it was assumed that the mirror magnetic field vanishes at the Earth, in which
case the $n$--$n'$ oscillation probability in vacuum after a time $t$ depends on the applied field
$\vect{B}$ as $P_{\vec{B}}(t) = \sin^2(\om t)/(\om \tau)^2$, where $ \om = \frac12 | \mu \vec{B} | =
(B/1\,{\rm mG}) \times 4.5\,{\rm s}^{-1}$, with $B = |\vec{B}|$ and $\mu = -6 \cdot 10^{-12}\,$eV/G
the neutron magnetic moment. Under this assumption the first limit was set on the $n$--$n'$ oscillation
time, $\tau > 1\,$s, using the beam monitoring data from the famous experiment~\cite{Baldo}, which
provided the strongest limit $\tau_{n\bar n} > 0.9 \times 10^8\,$s on the neutron-antineutron
oscillation~\cite{Kuzmin}.

In ultra-cold neutron (UCN) traps (see~\cite{dubbers} for a recent review on cold and ultra-cold
neutrons and their phenmenology) the $n$--$n'$ oscillation can be tested via anomalous magnetic
field dependent losses of neutrons. With a neutron flight time between wall collisions of the order
of $t\sim 0.1\,$s, the experimental sensitivity can reach $\tau\sim 500\,$s~\cite{Pokot}.
  Several
dedicated experiments \cite{Ban,Serebrov,Altarev,Bodek,Serebrov2} were performed by comparing the
UCN losses in \emph{large} ($B > 10\,$mG) and \emph{small} ($b < 1\,$mG) magnetic fields.
For \emph{small} fields one has $\om t < 1$\,so that $P_{\vec{b}} = (t/\tau)^2$, while for {\it
  large} fields one has $\om t \gg 1$ and oscillations are suppressed, $P_{\vec{B}} < (1/\tau \om)^2
\ll (t/\tau)^2$.  In this way, lower bounds on the oscillation time were obtained, which were
adopted by the Particle Data Group~\cite{PDG}.  The strongest bound, again under the
\emph{no-mirror-field} hypothesis, is $\tau > 414\,$s at 90\% CL~\cite{Serebrov,PDG}.

However, the above limits become invalid in the presence of a mirror matter or mirror magnetic field
~\cite{EPJ}.  In particular, in the background of both ordinary $\vect{B}$ and mirror $\vect{B}'$
magnetic fields the $n$--$n'$ oscillation is described by the Hamiltonian \footnote{The
  phenomenology of $n-n'$ oscillations in case of many ($\sim 10^{32}$) parallel sectors was
  discussed in ref. \cite{Redi}.  }
\beq{H-osc}
H_{nn'} 
= \mat{  \mu \vect{B} \bsig }{\dm} {\dm} {\mu \vect{B}'  \bsig  } ,
\ee
where $\bsig=(\sigma_x,\sigma_y,\sigma_z)$ are the Pauli matrices.  The probability of $n$--$n'$
transition after flight time $t$ was calculated in ref.~\cite{EPJ}.  It can be conveniently
presented as
\beq{P}
P_{\vec{B} }(t)  = \cP_{B}(t)  + \cD_{\vec{B}}(t)  
=  \cP_{B}(t)  + D_B(t) \cos\beta\,,  
\ee
where $\beta$ is the angle between the vectors $\vect{B}$ and  $\vect{B}'$ and 
\beqn{PD}
&&  \cP_{B}(t) =  
\frac{ \sin^2[(\om-\om')t] }{2  \tau^2 (\om -  \om^\prime)^2 } 
\, + \, \frac{ \sin^2[(\om+\om')t] }{2 \tau^2 (\om + \om')^2}, 
\nonumber  \\ 
&&   D_B(t)=  
\frac{ \sin^2[(\om-\om')t] }{2  \tau^2 (\om -  \om^\prime)^2 } 
\, - \, \frac{ \sin^2[(\om+\om')t] }{2 \tau^2 (\om + \om')^2} , 
\eeqn
with $ \om = \frac12 | \mu B|$ and  $ \om' = \frac12 | \mu B' |$.   
By reversing the magnetic field direction the probability becomes $P_{-\vec{B}}(t) = \cP_{B}(t) -
D_B(t) \cos\beta $. It is thus convenient to study the asymmetry $P_{\vec{B}} - P_{-\vec{B}} =
2D_{\vec{B}}\cos\beta$ in the neutron losses.

In this work we analyze in detail the data acquired in experiment~\cite{Serebrov2} and find a
dependence of the neutron losses on the magnetic field orientation, with more than $5\sigma$
deviation from the null hypothesis. This anomaly can not be explained by standard physics, but can
be interpreted in terms of $n$--$n'$ oscillations in the background of a mirror magnetic field.
Needless to say, the possible presence of the latter is striking in the light of mirror matter as
dark matter, with strong implications for its direct search and its possible accumulation in the
Earth.

\SEC{Experiment and data analysis.} The experiment \cite{Serebrov2} was carried out at the ILL,
Grenoble, using the well-known UCN facility PF2.
The trap of $190\,\ell$ volume capable of storing about half million neutrons was located inside a
shield screening the Earth magnetic field. A controlled magnetic field was then induced by a system
of solenoids.  Unfortunately, its value was not measured all over the trap and its exact profile was
not studied.  The reference magnetic field was evaluated approximately as $B \approx 0.2\,$G, but
due to possible inhomogeneities, its real value could have up to 25\% uncertainty.

Each measurement, taking about 10~min, consisted of three steps: filling of the trap during 130\,s
by {\it unpolarized} UCN through the basic neutron guide; closing of the entrance valve and storing
of the UCN in the trap for 300\,s; opening of the exit valves, counting the survived neutrons during
130\,s by two independent detectors.  The incident neutron flux during the filling was monitored by
another detector located in the neutron guide.  

The results of all measurements are reported in~\cite{Serebrov2}.  Here we concentrate on
measurements in {\it vertical} magnetic fields directed up ($+$) and down ($-$), which were
performed in three series.
In the first series \emph{small} ($b < 1$ mG) and \emph{large} ($B \simeq 0.2$\,G) magnetic fields
were used, repeating the sequences $\{\vect{b}|\vect{B}\} =
\{+\vect{b},+\vect{B},-\vect{B},-\vect{b}; -\vect{b},-\vect{B},+\vect{B},+\vect{b} \}$.
Unfortunately, the neutron flux was strongly unstable, counts randomly fluctuated and soon the
reactor was stopped for technical reasons.  Due to this, only a small part of the data records,
consisting of $\cN$=100 measurements for each of the $\pm \vect{B}$ and $\pm \vect{b}$
configurations, could be selected as acceptable for analysis.\footnote{Namely, three bands were
  selected in which the reactor power and the UCN flux were stable enough, with deviations no more
  than 10\% off the values of the normal functioning.}
In a second series, only the {\it large} magnetic field $B \simeq 0.2\,$G was employed, repeating 50
times the cycle $\{ \vect{B} \} =
\{-\vect{B},+\vect{B},+\vect{B},-\vect{B};+\vect{B},-\vect{B},-\vect{B},+\vect{B}\}$, for a total of
$\cN$=400 measurements in 72 hours of operation.  The next 24 hours were devoted to the calibration
tests in the UCN flow regime, totalling $\cN$=216 measurements (see later).  The experiment was
concluded by a third series of 16 cycles $\{\vect{2B} \}$ ($\cN$=128) under a magnetic field $2B
\simeq 0.4\,$G.

The neutron mean free-flight time between wall collisions and its variance were estimated via Monte
Carlo simulation~\cite{Serebrov,Serebrov2}.
For a storage time of $300\,$s one has $\langle t \rangle = t_\ff = 0.094\,$s and $ \langle t^2
\rangle - t_\ff^2 = \sigma_\ff^2 = 0.0036\,$s$^2$.
For estimating the mean oscillation probability $\ov{P}_{\vec{B}}=\ov{\cP}_B+\ov{\cD}_{\vec{B}}$,
the time dependent factors in (\ref{PD}) must be averaged over the UCN velocity distribution in
the trap.
The Monte Carlo simulated average concides with very good accuracy (percent) with the analytic
approximation $\langle \sin^2(\omega t) \rangle = S(\omega)= \frac12 \left[1 - \exp(-2\om^2
  \sigma_\ff^2) \cos(2 \om t_\ff) \right]$, that we adopt.  As a result, in the limit $\om t_\ff \ll
1$ we obtain $S(\omega) = \om^2 \langle t^2 \rangle$, while for $\om t_\ff \gg 1$ the oscillations
are averaged and $S(\om) = 1/2$.  In analyzing below the consequences for the mirror magnetic field
$B'$, the averages of the oscillating factors $\langle \sin^2\big[(\om\pm \om')t\big] \rangle =
S(\om\pm \om')$, might be safely set to $1/2$ unless $\om \approx \om'$.  In fact, the explicit form
of $S(\om- \om')$ is relevant only very close to the resonance, where $|B-B'| \sim 10^{-3}\,$G. 
In the resonance one has $\ov{\cP}_B,\ov{D}_B = \langle t^2\rangle /2\tau^2$.
Since $n$--$n'$ oscillation can take place not only during the 300\,s of UCN storage but also during
filling and emptying of the trap, the effective exposure time can be estimated as $t_\ast \approx
370\,$s~\cite{Serebrov2}.  Hence, for an overall amount of wall scatterings we take $n_\ast =
t_\ast/t_\ff \simeq 4000$.

\begin{table}[t]
  \small
  \def\arraystretch{1.1}
  \begin{tabular}{l |c|c}
    & $A_{\vec{B}}^{\rm det}(t_\ast) ~~ [ \times 10^{-4}]$  &  $A_{\vec{B}}^{\rm nor}(t_\ast) ~~ [\times  10^{-4}]$ \\ 
    \hline
    \hline
    $\{\vect{b} | \vect{B}\}$  
    &  ~ $2.03 \pm 2.69$  ~{\sl (1.45)}  &  ~ $4.04 \pm 3.05$  ~{\sl (1.04)}  \\
    & $-4.11\pm 3.80$ ~{\sl (1.36)} $^\dagger$  & $-3.23\pm 4.31$ ~{\sl (1.25)} $^\dagger$ \\ 
    \hline
    $\{\vect{B}\}$ 
    &  ~ $6.96 \pm 1.34$ ~{\sl (0.87)}  &   ~ $6.02 \pm 1.52$  ~{\sl (0.89)}    \\
    \hline
    $\{\vect{2B}\}$ 
    & $-0.26 \pm 2.40$ ~{\sl (1.77)}   & $-0.10 \pm 2.72$  ~{\sl (1.82)} 
  \end{tabular}
  \vspace*{-1ex}
  \caption{Results for $A_{\vec{B}}$ (and $E_B$, marked by $^\dagger$) by data fitting of three series, 
    their statistical errors and  the respective $\chi^2_{\rm dof}$  in parentheses.\label{Table}}%
  \vspace*{-1ex}
\end{table}

The raw data~\cite{Serebrov2} can be tested for magnetic field dependence of UCN losses, as a probe
for $n$--$n'$ oscillation.  In fact, if between the wall collisions the neutron oscillates into a
sterile state $n'$, then per each collision it can escape the trap with a mean probability
$\ov{P}_{\vec{B}}$.  The asymmetry in the magnetic field between the detector counts
$N_{\vec{B}}(t_\ast) \propto \exp(-n_\ast \ov{P}_{\vec{B}}) $ and $N_{-\vec{B}}(t_\ast) \propto
\exp(-n_\ast\ov{P}_{-\vec{B}} ) $, directly traces the difference between the probabilities
$\ov{P}_{\vec{B}} - \ov{P}_{-\vec{B}}= \ov{\cD}_B$ \cite{EPJ}:
\beq{AB}
A_{\vec{B}}^{\rm det}(t_\ast) = \frac{N_{-\vec{B}}(t_\ast) -
  N_{\vec{B}}(t_\ast)} {N_{-\vec{B}}(t_\ast)+N_{\vec{B}}(t_\ast)}
= n_\ast    \ov{D}_B  \cos\!\beta\,,
\ee 
where we assume $n_\ast \ov{\cD}_{\vec{B}} \ll 1$.  
Clearly, the neutron loss factors related to regular reasons, which are magnetic field independent,
cancel out from this ratio. These are the decay, the wall absorption or upscattering due to
collisions with the residual gas, etc.\footnote{As it was shown in ref. \cite{Kerbik}, 
the quantum mechanical corrections to $n-n'$ transition probability due to the finite size 
of the UCN traps are negligible.}
On the other hand, since $\ov{P}_{\vec{B}} + \ov{P}_{-\vec{B}} = 2\ov{\cP}_B $, the value
\beq{EB}
E_B^{\rm det}(t_\ast) = 
\frac{N_{\vec{b}}(t_\ast) +N_{-\vec{b}}(t_\ast)} {N_{\vec{B}}(t_\ast) +N_{-\vec{B}}(t_\ast)} -1 
= n_\ast (\ov{\cP}_B - \ov{\cP}_{b})  
\ee
should not depend on the magnetic field orientation.   

We compute then the values (\ref{AB}) and (\ref{EB}) by summing up the counts in two detectors,
$N=N_1+N_2$ (the individual counts $N_1$ and $N_2$ are used below for the stability check).  For
each detector we consider Poisson statistics, so that $\Delta N_{1,2} = \sqrt{N_{1,2}}$.  In
addition, we compute analogous asymmetries $A^{\rm mon}_{\vec{B}}$, $E^{\rm mon}_{B}$ for the
monitor counts $M_{\vec{B}}$ and $M_{-\vec{B}}$, and for the detector-to-monitor normalized ones
$A^{\rm nor}_{\vec{B}}$, $E^{\rm nor}_{B}$ using the ratios $(N/M)_{\vec{B}}$ and
$(N/M)_{-\vec{B}}$.

\begin{figure}[t]
  \centerline{\includegraphics[width=.97\columnwidth]{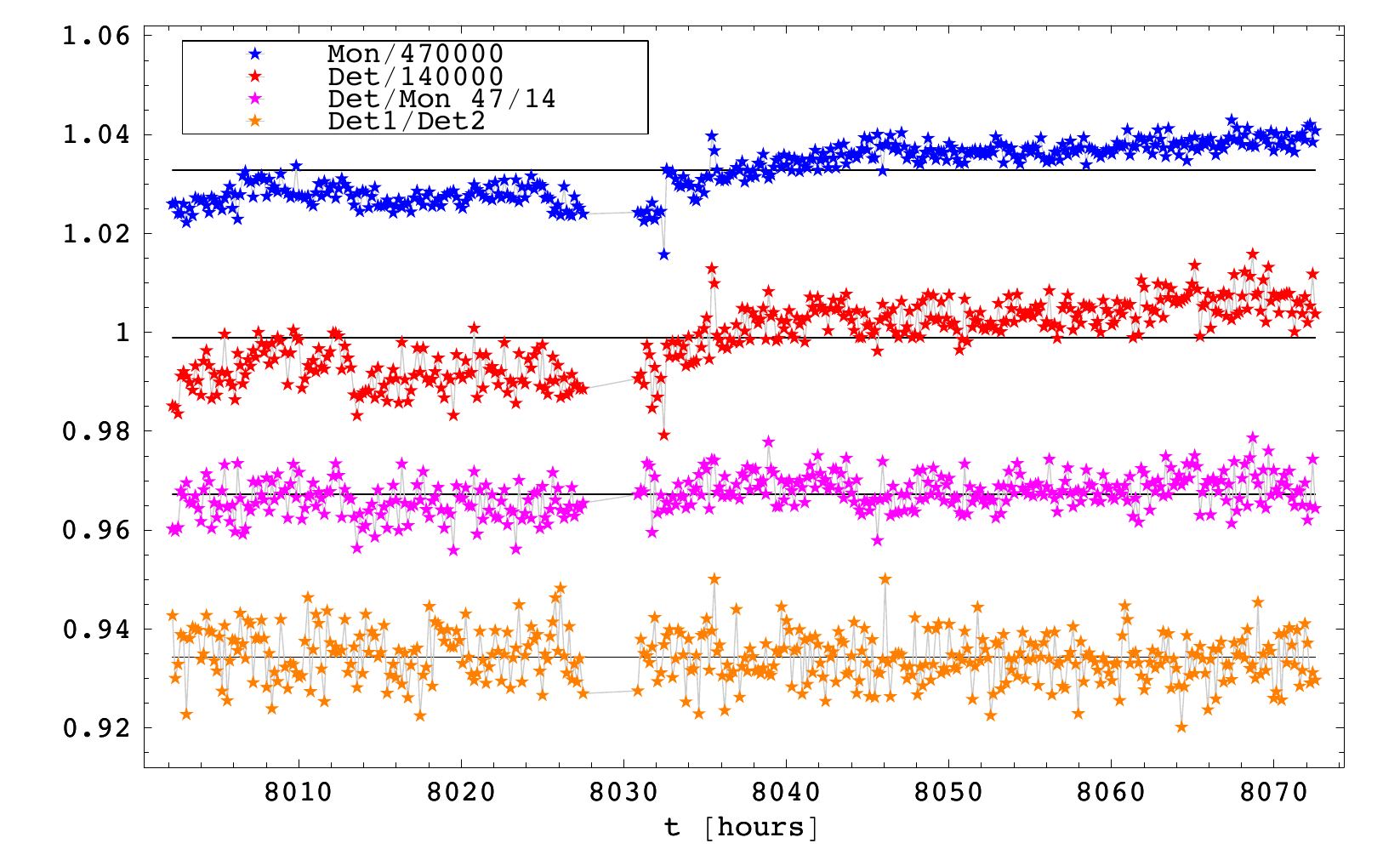}}%
  \centerline{\includegraphics[width=1.04\columnwidth]{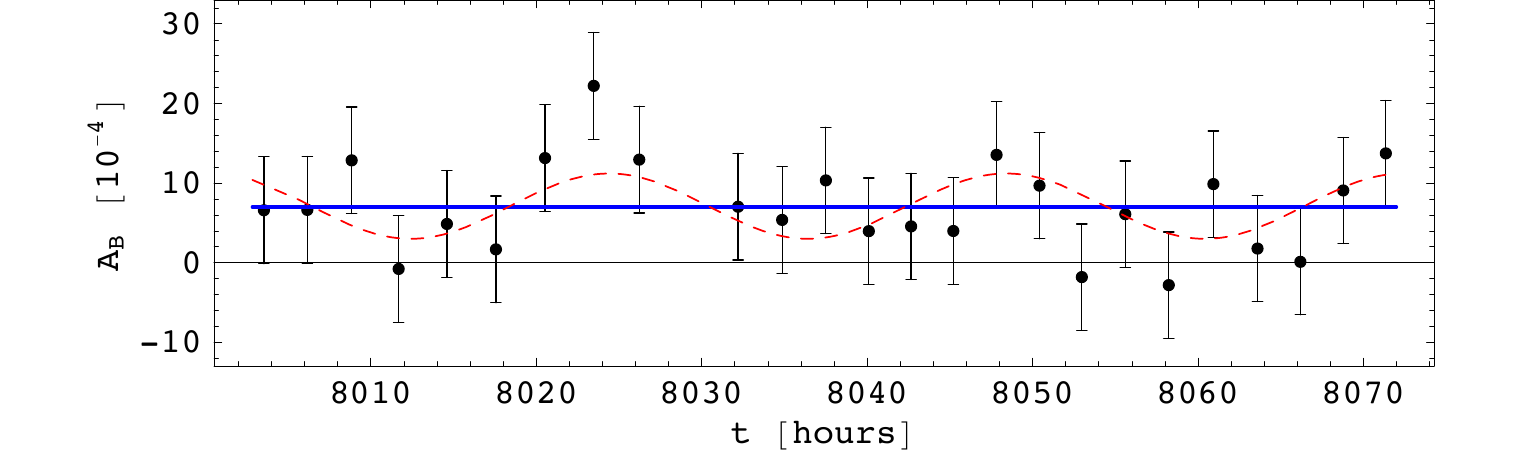}~~}%
  \vspace*{-1.5ex}
  \caption{The $\{\vect{B}\}$ series. \emph{Upper Panel}: from up to down, monitor counts $M$ and sum
    of detector counts $N=N_1+N_2$, normalized respectively to 470000 and 140000; then ratios $N/M
    (\times47/14$) and $N_1/N_2$.  \emph{Lower Panel}: $A^{\rm det}_{\vec{B}}$ binned by two
    $\{\vect{B}\}$ cycles (16 measurements), with the constant and periodic fits.}%
  \label{fig:30nov}%
  \vspace*{-1ex}
\end{figure}

The results are shown in Table \ref{Table}.  We see that the value of $A_{\vec{B}}^{\rm det}$, based
on $400$ measurements in $\{\vect{B}\}$ mode (see Fig.~\ref{fig:30nov}), has a $5.2\sigma$ deviation
from zero.\footnote{In ref.~\cite{Serebrov2} a somewhat different fitting procedure was adopted.
  The data were averaged between the $B$ and $2B$ magnetic fields and, as a result, a circa
  $3\sigma$ deviation was reported, which in our notation translates to $A^{\rm
    det}_{(\vec{B}+\vec{2B})} =(3.8 \pm 1.2) \times 10^{-4}$.  However, because the probability of
  $n$--$n'$ oscillation (\ref{PD}) depends resonantly on the magnetic field, one should not average
  between different field values.  After our communication, A.P.~ Serebrov and A.K.~Fomin reanalyzed
  the experimental records and confirmed the 5.2\,$\sigma$ anomaly in the $\{\vect{B}\}$ mode data.
  We thank them for this cross check. 
  For a joint proposal of new experimental series, to confirm
  definitely this anomaly or to exclude it, see~\cite{ILL}.}

Can this anomalous dependence on the magnetic field be induced by technical factors as e.g.\
fluctuation of the reactor power or unstable vacuum condition in the trap?  Fig.~\ref{fig:30nov}
shows that the detector counts $N$ had up to $2\%$ drift which is, however, well traced by the
monitor counts $M$: the constant fit of ratios $N/M$ gives $\chi^2_{\rm dof}= 1.55$. In addition,
individual counts in two detectors are perfectly synchronous: $N_1/N_2$ is constant with
$\chi^2_{\rm dof}= 0.98$.  In fact, the two detectors separately give $A^{\rm det1}_{\vec{B}}=(8.40
\pm 1.92)\times 10^{-4}$ ($\chi^2_{\rm dof}=0.88$) and $A^{\rm det2}_{\vec{B}}=(5.62 \pm 1.86)
\times 10^{-4}$ ($\chi^2_{\rm dof}=0.81$). It is important to note that since the measurements with
switching field were taken at consecutive times, a drift in the reactor flux (or changing vacuum
conditions or other factors that may affect the initial amount of neutrons in the trap) could
contaminate the asymmetry itself. However, the cycles $\{\vect{B}\}$ were configured to make the
asymmetry (\ref{AB}) insensitive to any slow drift.  Clearly, a linear drift is cancelled in each of
the measurement quartets $(-,+,+,-)$ and $(+,-,-,+)$, while the quadratic component is cancelled
between two consecutive quartets.  In fact, we fit $A_{\vec{B}}^{\rm det}$ as the average of
(\ref{AB}) in each complete $\{\vect{B}\}$ cycle (8 measurements), and obtain an excellent
$\chi^2_{\rm dof}=0.87$.

As a futher check, the anomaly cannot be eliminated as well by normalizing to the monitor counts: we
find a residual $4\sigma$ asymmetry also in $A_{\vec{B}}^{\rm nor}$ (see Table~\ref{Table}).  This
lower value is in agreement with the fact that this measure mildly underestimates the effect, at
first by statistical reasons: accounting for the monitor fluctuations $\Delta M = \sqrt{M}$ one
formally enlarges the errors; then, by dynamical reasons: during the 130\,s of filling time nearly
half of the neutrons counted by the monitor are neutrons that reenter the neutron guide back 
from the trap, where they could oscillate into $n'$ being exposed to the magnetic field.  
The UCN diffusion time in the trap when the entrance valve is open is estimated as 
$t_{\rm dif} \simeq 60$\,s.
Hence, the monitor asymmetry $A^{\rm mon}_{\vec{B}}$ is expected to be one order of magnitude less
than $A^{\rm det}_{\vec{B}}$.  In fact, analyzing the monitor data we get $A^{\rm
  mon}_{\vec{B}}=(0.96 \pm 0.72)\times 10^{-4}$ ($\chi^2_{\rm dof}= 0.90$).

Finally, a series of calibration measurements were performed in order to check for possible
systematic effects that could make the neutron counts sensitive to the magnetic field orientation,
as for instance an influence of the alternating solenoid current on the counting electronics.
Measurements were performed with high statistics in $\{\vect{B}\}$ mode, with data taken in
continuous flow regime, i.e.\ with entrance and exit valves of the trap open during 200\,s of
counting simultaneously with the two detectors and the monitor.  With valves open, the effective
diffusion time of the UCN in the trap is estimated via MC simulations as $t^{\rm flow}_\ast \simeq
20\,$s.  Coherently, these counts show no systematic effects: we find $A_{\vec{B}}^{\rm det} =(0.01
\pm 0.39) \times 10^{-4}$ ($\chi^2_{\rm dof} =1.23$) and $A_{\vec{B}}^{\rm mon}=(0.22 \pm 0.78)
\times 10^{-4}$ ($\chi^2_{\rm dof} =1.16$).  The counts of the two detectors were stable: the ratio
$N_1/N_2$ is fitted by a constant with $\chi^2_{\rm dof} =0.98$.

\begin{figure*}[t]
  \centerline{\includegraphics[width=.99\textwidth]{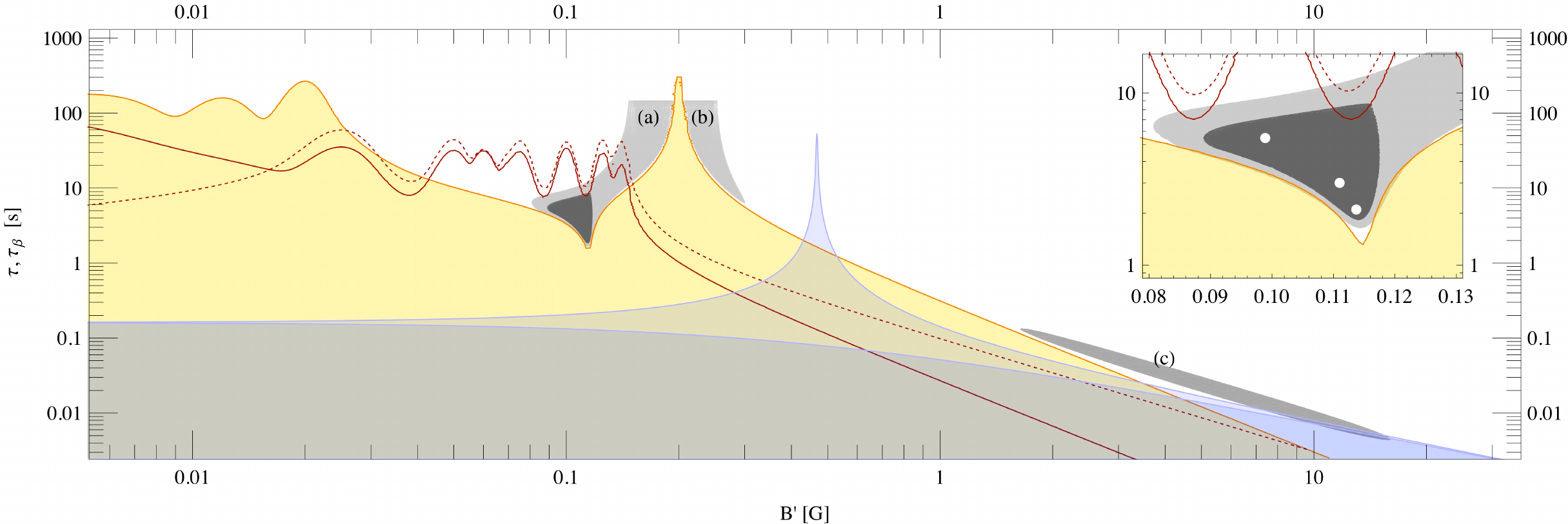}}
  \vspace*{-1ex}
  \caption{Global fit in the $B'$-$\tau$, $\tau_\beta$ plane.  The positive result (anomaly)
    corresponds to the gray-shaded areas, which show the parameter space allowed at $90\%$\,CL
    (darker) and $99\%$\,CL (lighter) by the global fit of non-zero $\ov{D}_B$, eq.~(\ref{DB}), with
    magnetic field marginalized over the uncertain range $B=0.15-0.25$\,G (the zoomed inset displays
    the best fit points assuming a constant field $B=0.15,\, 0.20, \, 0.25$, left to right). For
    comparison, available constraints from earlier measurements are also shown: the yellow-shaded
    area in the background is excluded at $99\%$\,CL by the measurements of $E_B$ from
    refs.~\cite{Serebrov,Serebrov2}; the region of $\tau$ ($\tau_\beta$) below the {\it wavy} solid
    (dotted) curves are disfavored by the measurements of refs.~\cite{Ban,Bodek,Altarev} (not
    included in the fit).  Interestingly, the data of ref.~\cite{Altarev} for $E_B$ and $A_B$ also
    imply a best fit value $B'=0.11\,$G, with $\tau = 14\,$s and $\tau_\beta = 20\,$s respectively.
    The blue-shaded area peaked at $B'=0.5\,$G is excluded by measurements in the Earth magnetic
    field, illustrated for $B'$ and $B_{\rm Earth}$ parallel (lighter blue) and antiparallel (darker
    blue).}
  \label{fig:plot}%
  \vspace*{-1ex}
\end{figure*}

\SEC{Interpretation of the results.} Let us now analyze the obtained results 
in the light of $n$--$n'$
oscillations.  Using (\ref{AB}) and (\ref{EB}), the values shown in Table~\ref{Table} translate into
\bea
\ov{D}_B \cos\beta =  (1.60 \pm 0.32) \times 10^{-7}  ~~  \label{DB}  \\
\ov{\cP}_B - \ov{\cP}_b 
= -(1.03 \pm 1.11) \times 10^{-7} \label{PB}\\
\ov{D}_{2B} \cos\beta =-(0.06 \pm 0.80) \times 10^{-7}   \label{D2B} 
\eea
where we have conservatively taken $A_{\vec{B}}^{\rm det} =(6.40 \pm 1.26) \times 10^{-4}$, by
averaging the results of $\{\vect{B}\}$ and $\{\vect{b}|\vect{B}\}$) cycles.

Eqs.~(\ref{P}) and (\ref{PD}) show that in the presence of strong enough mirror field, $B' \gg
10$\,mG, the values of $\cP_B$ and $D_B$ have peculiar dependence on the experimental magnetic
field, so the above results can be used to put constraints in the plane $(B', \tau)$ or $(B',
\tau_\beta=\tau|\cos\beta|^{-1/2})$.

Eq.~(\ref{DB}), for a given $B$, gives a correlation between $B'$ and $\tau_\beta$.  We perform a
2-parameter fit in this plane, and find the preferred regions which are depicted as gray areas in
Fig.~\ref{fig:plot}.  Since the homogeneity of the vertical field $\vect{B}$ was not precisely
controlled in this experiment, and its effective value averaged over the trap could vary between
$B=0.15 - 0.25\,$G, we consider that $(B/0.2\,{\rm G}) = 1 \pm 0.25$ and marginalize over this
range. 
The global fit also includes the constraint from~(\ref{PB}), conservatively referring to
the case $\cos\beta=1$, as well the limits on $\tau$ from experiments with horizontal magnetic
field~\cite{Serebrov,Serebrov2} and the limit on the neutron losses in the Earth magnetic
field~\cite{Earth}.
These latter limits are also explicitly depicted, respectively as the yellow area peaked at 0.2\,G
and the blue area peaked at 0.5\,G.  The horizontal field measurements of ref.~\cite{Serebrov2}
(with $B=0.2$\,G) imply $\cP_B - \cP_b=-(3.60\pm 1.95) \times 10^{-8}$. For $B'\gg1\,$G this gives
the lower limit $\tau > 0.28\,{\rm s} \times (1\,{\rm G}/B')^2$.  The measurements of neutron losses
in the Earth magnetic field ($B\approx 0.5\,$G) yield roughly $P_B < 2\times
10^{-6}$~\cite{Earth}. For $B' \gg 1\,$G it gives the limit $\tau > 0.1\,{\rm s} \times (1\,{\rm
  G}/B')$.

As one can see from Fig.~\ref{fig:plot} the positive  asymmetry~(\ref{DB}) along with
the constraint (\ref{PB}) and the limits from horizontal field measurements \cite{Serebrov,
  Serebrov2}, restrict the parameter space to three regions marked as (a), (b) and (c).

The $\ov{D}_B$ asymmetry and $\ov{\Delta}_B$ imply that the preferred region is (a), where the
mirror magnetic field $B' =0.09$~to~$0.12$\,G at 90\% CL, and the $n$-$n'$ oscillation time is in the
range $2$ to $10$\,s.  The region is considerably enlarged by the $B$ magnetic field uncertainty
which is marginalized in the fit.  The best fit point, visible in the figure inset, is relative to
$B=0.2$\,G and corresponds to $B'=0.11$\,G, $\tau_\beta=3$\,s.

At 99\% CL the region becomes larger and also region (b)\ beyond the 0.2\,G resonance (of the
horizontal-field measurements) becomes allowed. The region extends up to $B'\simeq 0.3$\,G,
therefore we conclude that at 99\% CL the mirror magnetic field is constrained in the range
$0.08\,{\rm G}<B'<0.3$\,G.

We note finally that at larger $B'$ the horizontal-field measurements do not constrain the positive
result of $\ov{D}_B$ and a third region (c)\ is allowed, extending from $B'=1.5$\,G to 15\,G where
the Earth-field constraint becomes dominant, with oscillation time in the range $0.15\,{\rm s}>
\tau_\beta > 0.005$\,s.  This region has however higher minimum $\chi^2$ and in addition it is
disfavored by the constraint (\ref{D2B}).

\medskip

The positive result that emerged from the fit points to a nonzero mirror magnetic field at the
Earth. Let us then comment whether this is plausible. If mirror particles represent dark matter,
they must present in the Galaxy along with the normal matter.  If by chance the solar system is
traveling across a giant molecular cloud extended over several parsecs, there may exist a mirror
field $\vect{B}'$, with $B' \sim 10$ to $100\,$mG.  Then, since the experimental field $\vect{B}$
rotates together with the Earth, the angle $\beta$ between $\vect{B}$ and $\vect{B}'$ and thus
$P_{\vec{B}}$ would show a periodic time dependence with period of sidereal day T=23.94\,h.
On the other hand, if there exist strong enough interactions between ordinary and mirror particles,
e.g.\ due photon--mirror photon kinetic mixing~\cite{Holdom}, then the Earth can capture a
significant amount of mirror matter.\footnote{According to ref.~\cite{IV}, the geophysical data 
on the Earth mass, moment of inertia, normal mode frequencies etc.\ allow 
the presence of mirror matter in the Earth with mass fraction up to  $4\times 10^{-3}$.   } 
The natural capture asymmetry due to the Earth rotation would also give rise to circular currents
that could induce a mirror magnetic field up to several Gauss~\cite{EPJ}.  If the captured mirror
matter forms a compact body rotating sinchronously with the Earth, then $\beta$ would not vary in
time.  However, if it forms an extended halo around the Earth with a differential rotation, the
mirror field $\vect{B}'$ and hence $P_{\vec{B}}$ may have more complex time variations.

Interestingly, the data of series $\{\vect{B}\}$ hint to a periodic time-dependence, consistent with
sidereal day period (see Fig.~\ref{fig:30nov}).  Fitting the up-down asymmetry as $A^{\rm
  det}_{\vec{B}} = C + V \cos \big[\frac{ 2\pi}{\rT}(\rt-\rt_0) \big]$ (4 parameters) we obtain $C =
(7.09 \pm 1.26)\times 10^{-4}$, $V = (4.10\pm 1.71) \times 10^{-4}$, $\rT= 24.0 \pm 1.8\,$h and
$\rt_0=8000.4\pm 1.8\,$h, with $\chi^2_{\rm dof} =0.82$.
(Asymmetries in both detectors are consistent with such periodicity.)  Clearly, since the constant
fit already has a very good $\chi^2$, its further improvement with the periodic fit is not very
significative, and testing the time dependence requires more statistics.  To our regret, the data in
$\{\vect{b}|\vect{B}\}$ and $\{2\vect{B}\}$ were not broad and stable enough for a reliable
time-dependent analysis.

%
\vspace*{1ex}

\SEC{Summary.} The phenomenon of $n$--$n'$ oscillation is particularly attractive, especially in the
light of our findings, which clearly call for future experiments with higher precision. In
particular, using the same $190\,\ell$ UCN chamber with $t_{\rm f} \simeq 0.1\,$s as in the
experiments~\cite{Serebrov,Serebrov2} at the ILL PF2 EDM facility, these oscillations can be tested
under properly controlled magnetic field profiles~\cite{ILL}.  By tuning the magnetic field to the
resonance value $B = B'$ with a precision of 1\,mG, the probability of $n$--$n'$ transition can be
increases up to $\cP_{\rm res},D_{\rm res} \simeq (t_{\rm f}/\tau)^2$, i.e.\ $\sim10^{-3}$ for $\tau
= 3$\,s. Then the neutron losses would be very sizable, $A_B\sim 0.1$, and also neutron regeneration
$n\to n' \to n$ and resonant corrections to the neutron spin-precession~\cite{EPJ} could be
optimally tested.  If the DUSEL project~\cite{DUSEL} will be realized, the neutron flight time could
be increased up to few seconds which would allow to test the $n$--$n'$ oscillation in an exhaustive
way.

Concluding, the experimental data~\cite{Serebrov2} indicate that the neutron losses in the UCN trap
in magnetic field $B\simeq 0.2$\,G depend on the magnetic field direction, showing an anomaly about
$5\sigma$ deviated from the null hypothesis, which can not be interpreted by standard physics.  If
this anomaly will be confirmed by future experiments, it can be explained by neutron oscillations
into mirror neutrons, in the presence of a mirror magnetic field $B'\sim 0.1\,$G.  Such a discovery
would shed light also on fundamental physical problems as the nature of dark matter, primordial
baryogenesis, stability of neutron stars~\cite{BB-nn} and many other astrophysical issues as e.g.\
the origin of the pre-GZK cutoff in the cosmic spectrum \cite{UHECR}.  In addition, the underlying
physics at the scale $M\sim 10$~TeV could be testable at the LHC. 
The discovery of a parallel world via $n$--$n'$ oscillation 
and of a mirror magnetic background at the Earth,
striking in itself, would give crucial information on the accumulation the of dark matter in the
solar system and in the Earth, due to its interaction with normal matter, with far reaching
implications for physics of the sun and even for geophysics. 

\vspace*{1ex}

\SEC{Acknowledgements.}  We thank A.~Serebrov and PNPI-ILL collaboration for providing the data
records of the experiment~\cite{Serebrov2} and complete technical information.  We also thank
Yu.\ Kamyshkov and D.V. Naumov for discussions and useful suggestions, and D.V. Naumov also for
Monte-Carlo simulations for checking the effects expected from the neutron intensity drift.  
The work was supported in part by the Italian National grant PRIN2008 \emph{Astroparticle Physics}
and in part by the RF Science Ministry Grant No. 02.740.11.5220.

\end{document}